# Observation of the Einstein–de Haas Effect in a Bose–Einstein condensate

Hiroki Matsui[1], Yuki Miyazawa[1], Ryoto Goto[2], Chihiro Nakano[2], Yuki Kawaguchi[3,4], Masahito Ueda[5], Mikio Kozuma[1,2]*

[1]*Institute of Integrated Research, Institute of Science Tokyo; Tokyo, Japan.*
[2]*Department of Physics, Institute of Science Tokyo; Tokyo, Japan.*
[3]*Department of Applied Physics, Nagoya University; Nagoya, Japan.*
[4]*Research Center for Crystalline Materials Engineering, Nagoya University; Nagoya, Japan.*
[5]*Department of Physics, University of Tokyo; Tokyo, Japan.*

The Einstein–de Haas effect is a phenomenon in which angular momentum is transferred from microscopic spins to mechanical rotation of a macroscopic rigid body. Here, we report the first observation of the Einstein–de Haas effect in a spinor-dipolar Bose–Einstein condensate where quantized vortices emerge in depolarized spinor components through coherent angular-momentum transfer from microscopic atomic spins to macroscopic quantized circulation. Experimental results clearly show that the spherical symmetry of the condensate is dynamically broken into the axisymmetry by an intrinsic magnetic dipole-dipole interaction.

Experimental vindication of angular momentum transfer between microscopic spin and macroscopic mechanical rotation is a century-old problem in physics. A close link between magnetization and angular momentum (*1*) was experimentally demonstrated by Einstein and de Haas – a phenomenon known as the Einstein–de Haas (EdH) effect (*2*). The observed magnetization was later identified with electron spins, indicating that the total angular momentum of spin and mechanical rotation is conserved (*3*). The quest for understanding the microscopic mechanism of the EdH effect continues up until today, prompted by the observation of ultrafast spin dynamics induced by pulse lasers (*4–7*). Recent studies have demonstrated that chiral phonons can mediate angular-momentum transfer between spins and mechanical rotation of rigid bodies (*8–13*). However, the emergence of rigid-body rotation associated with spin relaxation is conventionally recognized on the basis of the law of conservation of the total angular momentum. It is of fundamental importance to experimentally test whether such macroscopic transfer of angular momentum can occur coherently. Quantum-mechanical description of this mechanism, including the underlying coupling mechanism for the mediator, is challenging due to the necessity of analyzing a finite-sized quantum many-body system (*14*).

Spinor-dipolar Bose–Einstein condensates (BECs) of dilute atomic gases offer a unique platform for investigating the EdH effect where a rigid body is replaced by a quantum condensate whose orbital angular momentum manifests itself in the form of quantized vortices in macroscopic wave functions (*15, 16*). Moreover, the transfer of angular momentum from spin to orbital degrees of freedom occurs spontaneously due to an intrinsic magnetic dipole-dipole interaction (MDDI) between atomic spins (Fig. 1A). So far, the exotic phenomena induced by the MDDI have been investigated using spin-polarized gases, focusing on their long-ranged and anisotropic nature (*17, 18*). The experimental realization of a spinor-dipolar BEC presents an opportunity to explore yet another feature of the MDDI, spin-orbit coupling. The EdH effect is a striking manifestation of this feature (*19–23*). Furthermore, the spin-orbit coupling stabilizes a circulating ground state by developing spin textures accompanied by spontaneous breaking of the chiral symmetry (*24, 25*).

Observing such macroscopic angular momentum conversion via the MDDI presents an experimental challenge due to the difficulty in reducing an ambient magnetic field down to below the magnetic dipole field. In this study, we report the first successful observation of the EdH effect in an optically confined BEC of neutral europium (Eu) atoms (*26*) whose electronic ground state is highly symmetric and possesses a large magnetic dipole moment of $7\mu_{\mathrm{B}}$. A gaseous BEC was prepared with the atomic spins initially aligned along the direction of an applied magnetic field. After the gas was temporarily held in a sufficiently weak magnetic field, we observed population transfer from the initially prepared magnetic sublevel to the other sublevels, which exhibited spatial distributions with ring-shaped structures, indicating creation of quantized vortices. Furthermore, the phase windings along the ring observed in atomic interferometry indicated that angular momentum was coherently transferred from the atomic spins to the quantized orbital angular momentum. We also performed numerical simulations which reproduced these observations, supporting the EdH effect induced by the MDDI.

## Preparation of atomic gas

Our experiment starts with a BEC of neutral $^{151}$Eu atoms in the hyperfine ground state with a spin of $F$=6. The BEC of a spin-6 Bose gas is represented by the macroscopic spinor wavefunction $\Psi(\mathbf{r}) = \sqrt{n(\mathbf{r})}\,\zeta(\mathbf{r})$, where $n$ is the density of atoms, and $\zeta = (\zeta_{+6}, \zeta_{+5}, \dots, \zeta_{-6})^{\mathrm{T}}$ is the normalized spinor, with $\zeta^\dagger\zeta = 1$. The gas is confined to an optical dipole trap whose potential minimum is nearly spherically symmetric and is characterized by a depth of 1.3 μK and oscillation frequencies of $(\omega_x, \omega_y, \omega_z)/(2\pi) = (110, 110, 130)$ Hz, where the *z*-axis is oriented vertically. The trap typically contained approximately $5\times10^4$ atoms, with a negligible thermal component. According to our numerical simulation, the magnitude of the magnetic dipole field was estimated to reach up to 8 nT. Thus, a magnetic field of 1 μT was applied to set the initial direction of the atomic spins. We take the quantization axis for the spinor $\zeta$ along this initial direction.

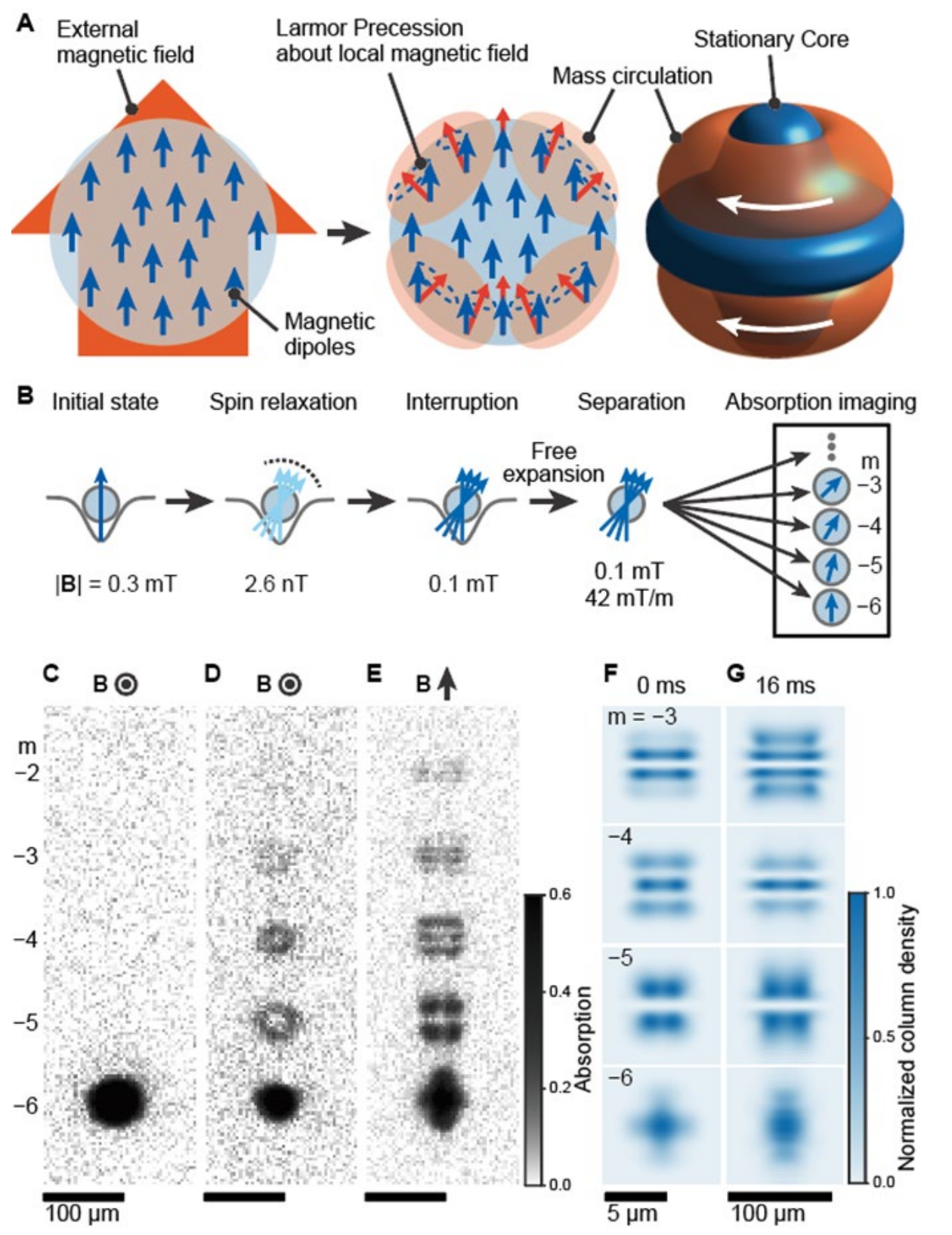

Fig. 1. Mechanism and observation of spin relaxation. (A) In the presence of an external magnetic field, atomic spins align parallel to an external magnetic field to minimize the Zeeman energy of the corresponding magnetic dipoles in the magnetic field. Upon removal of the magnetic field, the atomic spins undergo the Larmor precession about local magnetic fields, resulting in spin relaxation and mass circulation. (B) Spin relaxation proceeds in a weak magnetic field of 2.6 nT. The dynamics are then suppressed by the application of an external magnetic field of 0.1 mT. Subsequently, the gas is released from the trap and imaged after being spatially separated into spinor components. (C to E) Single-shot absorption image of the gas with spins initially aligned along the imaging axis (C and D) and the vertical direction (E) after a free fall for 16 ms. The gases are held in magnetic fields of 1.0 μT (C) and 2.6 nT (D and E) for a duration of 5 ms. (F and G) Simulated column density of the gases before (F) and after (G) free expansion of 16 ms in total. The color scales are normalized for the respective spinor components.

## Observation of spin relaxation

To demonstrate the EdH effect, we first observed spin relaxation in the BEC. The strength of the external magnetic field applied to the BEC was momentarily adjusted to a value at which the spin relaxation could occur, and subsequently to 0.10 mT to interrupt the spin relaxation. Throughout this procedure, the direction of the external magnetic field was maintained along the initial direction. The gas was then released from the trap and, 2.7 ms later, subjected to a magnetic field gradient of 42 mT/m for a duration of 6 ms. This procedure spatially separated the gas into the spinor components $\Psi_m(\mathbf{r}) = \sqrt{n(\mathbf{r})}\,\zeta_m(\mathbf{r})$ via the Stern–Gerlach effect (*27*) (Fig. 1B). Following a free fall of 16 ms after the release, the atoms were imaged using a standard absorption imaging technique. As this observation technique is destructive, a single experimental run, lasting 8 s in total, yields a single image.

The single-shot images shown in Fig. 1, C and D, and Fig. 1E were obtained for gases whose initial atomic spins were oriented along and perpendicular to the horizontal imaging axis, respectively. Each portion of the shadows in the images represents the column density of the corresponding spinor components $|\Psi_m(\mathbf{r})|^2$ after the free fall. In contrast to the gas held in a strong magnetic field of 1.0 μT (Fig. 1C), which exhibits a flawless round profile exclusively in the $m = -6$ component, gases held in a weak magnetic field of 2.6 nT (Fig. 1, D and E) reveal significant populations in the $m > -6$ components. This clearly indicates that the gas has undergone spin relaxation under the weak magnetic field. The value of 2.6 nT may be subject to a systematic error of up to 1.0 nT and has been chosen to ensure that the magnetic field is oriented along the intended direction under limited magnetic field stability (Materials and Methods). Furthermore, the $m = -5$ and $-4$ components comprise multiple parts, as shown in Fig. 1E, and clearly exhibit voids in the central portion of their spatial distribution (Fig. 1D). Given the nearly spherical shape of the trap potential, the two images (Fig. 1, D and E) can be interpreted as two orthogonal views of the spinor components. Consequently, the $m = -5$ and $-4$ components consist of two and three rings, respectively. Our numerical simulations revealed that the spinor components retain their spatial profile during free fall (Fig. 1, F and G) and excellently reproduce the experimental results shown in Fig. 1E. Therefore, the experimentally observed structures correspond to a magnification of the spatial distribution in the trap rather than interference fringes after free fall.

## Spin relaxation dynamics

The dynamical aspects of the spin relaxation were investigated by taking images of the atoms for varying hold times in a magnetic field of 2.6 nT. The results exhibited few shot-to-shot fluctuations, indicating that each image represents one moment in the overall dynamics. The typical results are shown in Fig. 2A. Spin relaxation progressed rapidly as soon as atoms were placed in the weak magnetic field, resulting in significant populations in the $m \leq -2$ components after 4 ms. Subsequently, the $m \geq -4$ components lost their distinctive patterns as atoms disappeared, whereas the $m = -5$ and $-6$ components retained two- and one-part structures, respectively.

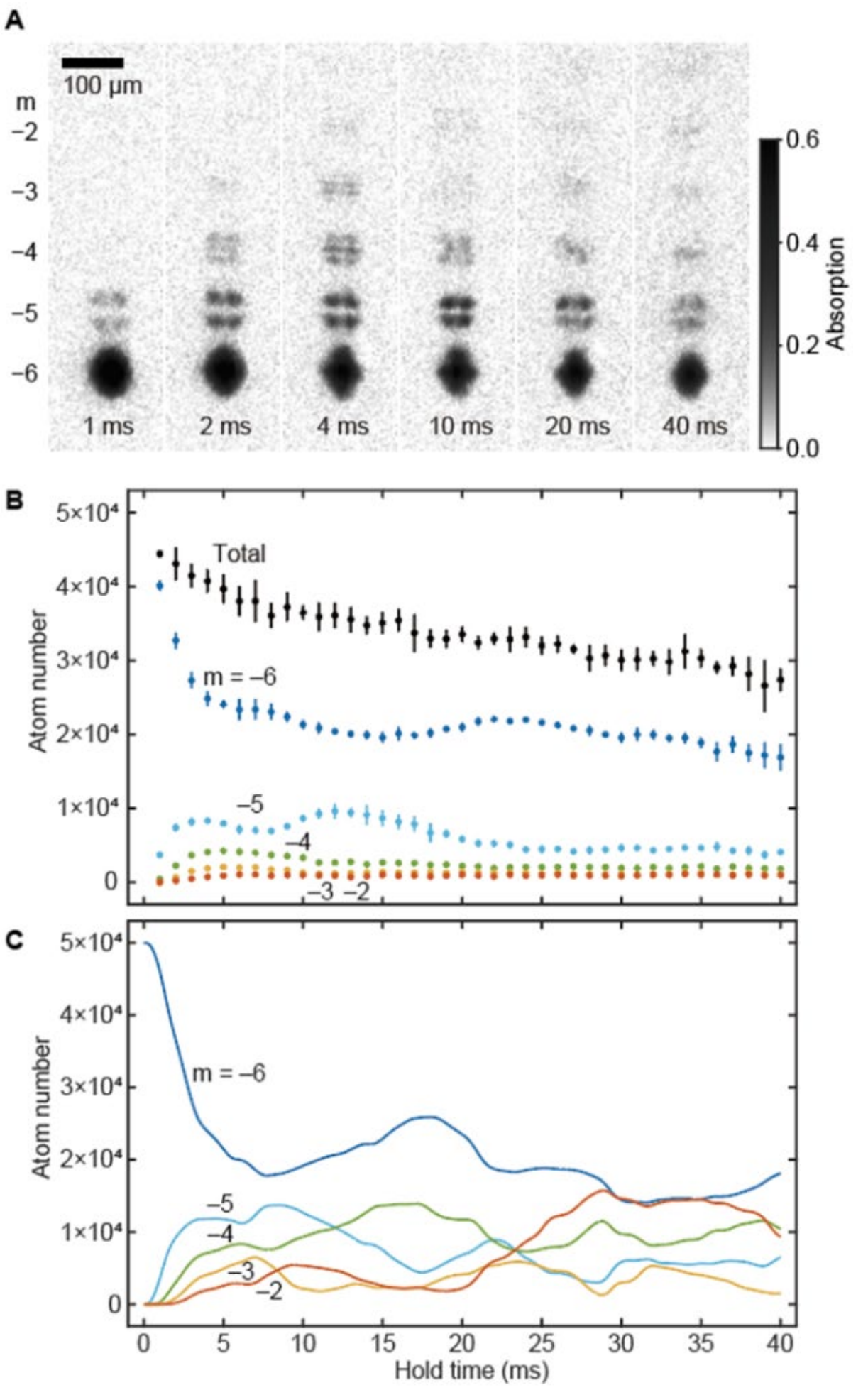

Fig. 2. Spin relaxation dynamics. (A) Single-shot absorption images of gases for various hold times in a magnetic field of 2.6 nT. (B) Dynamics of the populations in the spinor components held in a magnetic field of 2.6 nT. The populations in the m = –6,...,–2 components are shown (the m≥ –1 components are omitted for clarity). The total populations in all the components are also provided. Each data point is the average of four experimental iterations, and the error bars denote the standard deviations. (C) Numerical results for the in-situ populations corresponding to (B). The total population in (C) remains constant since collisional losses are not included in our model.

The populations of each spinor component were determined from the absorption images (*28*) and are shown in Fig. 2B, along with their total sums. The total population decreased by approximately 40% over 40 ms, which is in stark contrast to the 2% loss observed for the same duration under a magnetic field of 1.0 μT, where no spin relaxation occurred. This indicates the presence of spin-dependent loss mechanisms, which may account for the discrepancy in the $m \geq -4$ components between our loss-free numerical simulation (Fig. 2C) and experimental results (Fig. 2B). In contrast, the two results reasonably agreed for the $m \leq -5$ components. Specifically, the $m = -6$ component decreased in population shortly after the gas was subjected to a weak magnetic field, whereas the $m = -5$ component increased in population. Eventually, the $m = -6$ component partially recovered its population, reaching a maximum at 22 ms in the experiment, and subsequently decayed together with the $m = -5$ component. The extended period to reach the maximum compared with that in the numerical simulation can be attributed to the weak magnetic dipole fields due to the reduced populations. If the duration is considered to be a period of the Larmor precession of atomic spins around a local magnetic field, its frequency of 45 Hz corresponds to a magnetic field of 2.7 nT, which falls within the range of numerically simulated magnetic dipole fields (< 8 nT) produced by the atomic gas.

## Angular momentum transfer

Angular momentum transfer via the MDDI can be understood by considering the initial stage of the dynamics. The atomic spins are initially fully polarized along the *z*-axis, i.e., the $m \geq -5$ components are absent: $\Psi_m(\mathbf{r}) = \delta_{m,-6}\sqrt{n(\mathbf{r})}$ ($\delta_{i,j}$ is the Kronecker delta). At this stage, population transfer from the $m = -6$ component to the $m = -5$ component is governed only by the MDDI even in the presence of the spin-dependent short-range interaction: The population transfer comes from the spins' Larmor precession about the local magnetic field. Here, the relative phase between the wavefunctions of the $m = -6$ and $-5$ components corresponds to the direction of the transverse magnetization. Considering the rotational symmetry around the *z*-axis and the odd inversion symmetry about the $z = 0$ plane of the transverse component of the magnetic dipole field (Fig. 1A) in the cylindrical coordinates $(\rho, \phi, z)$, the spatial dependence of the growing $m = -5$ component $\Psi_{-5}(\mathbf{r})$ is given by $-\mathrm{i}z\rho\exp(-\mathrm{i}\phi)$ times a positive function of $\rho$ and $|z|$ (see Supplementary Material for details). Accordingly, the growing $m = -5$ component contains a line node along the $\rho = 0$ axis and a plane node on the *z* = 0 plane, as well as rotational symmetry of the norm, which corresponds to the double-ring structure. The phase difference of $\pi$ across the $z = 0$ plane may contribute to the stabilization of the double-ring structure for a prolonged period (Fig. 2A). Furthermore, the $\phi$ dependence of the growing $m = -5$ component $\Psi_{-5}(\mathbf{r})$ shows that a phase winding of $2\pi$ around the $z$-axis, i.e., mass circulation, is simultaneously imprinted onto the wavefunction as population transfer between the spinor components takes place. In other words, an angular momentum of $\hbar$ is transferred from the atomic spin to the orbital degrees of freedom, which is the hallmark of the EdH effect. As a result of local spin-gauge symmetry (*29*), the resulting spin $\mathbf{f}(\mathbf{r})$ tilts in the tangential direction, forming a helical spin texture.

**Observation of mass circulation**

To demonstrate the EdH effect, we next discuss mass circulation in the BEC, which serves to compensate for the loss of angular momentum from the atomic spins. We extracted information on the phase of the gas via matter-wave interferometry, as shown in Fig. 3A. A BEC initially spin-polarized along the vertical direction was subjected to a weak magnetic field and subsequently released from the trap following an interruption of spin relaxation, as in the aforementioned experiments. The free-falling gas was coherently split into halves by each of two consecutive Bragg scattering processes at moving optical crystals (*30*, *31*), thereby forming an open-type atom interferometer (*32–34*). The outputs were then separated into the spinor components by a brief application of a magnetic field gradient, and the resulting interference fringes were imaged along the vertical direction.

Figure 3B shows a typical single-shot image of the interferometer outputs for an atomic gas subjected to a magnetic field of 2.6 nT for a duration of 5 ms. The interference fringes, which are barely visible due to random noise, manifests itself by averaging the images over 116 iterations, as shown in Fig. 3C. Interference fringes are clearly visible for the $m \leq -4$ components, as shown in Fig. 3D, indicating that the fringes emerge stably at the same locations throughout the iterations. As is usually observed in interference between two BECs without vortices, parallel stripes are found in the fringes for the $m = -6$ components (*35*). In contrast, the fringes for the $m = -5$ component exhibit gently curved stripes with an interrupted stripe at each edge of the central void. This represents typical fringes observed in interference between two displaced identical BECs with vortices (*33*, *34*, *36*) and clearly shows the presence of a vortex with a phase winding of $2\pi$. Notably, the theoretically predicted $\pi$ phase difference between the two rings in the $m = -5$ component does not cancel the interference pattern, as interference occurs within the respective rings. In the $m = -4$ component, the stripes in the void region are not visible due to the limited number of atoms. However, the stripes around the void are not aligned with each other across the void, which is indicative of the existence of vortices with a non-zero winding number. This also holds for the $m = -5$ component in the image obtained for the gas held for an extended duration of 40 ms (Fig. 3, E and F). The direction of circulation can be determined from the shift in the stripes across the central void. As shown in Fig. 3, D and F, it is clockwise. This is consistent with the fact that the angular momentum is transferred from spins to the orbital angular momentum as a consequence of the conservation of total angular momentum. Therefore, we conclude that the observed spinor dynamics vindicates the EdH effect.

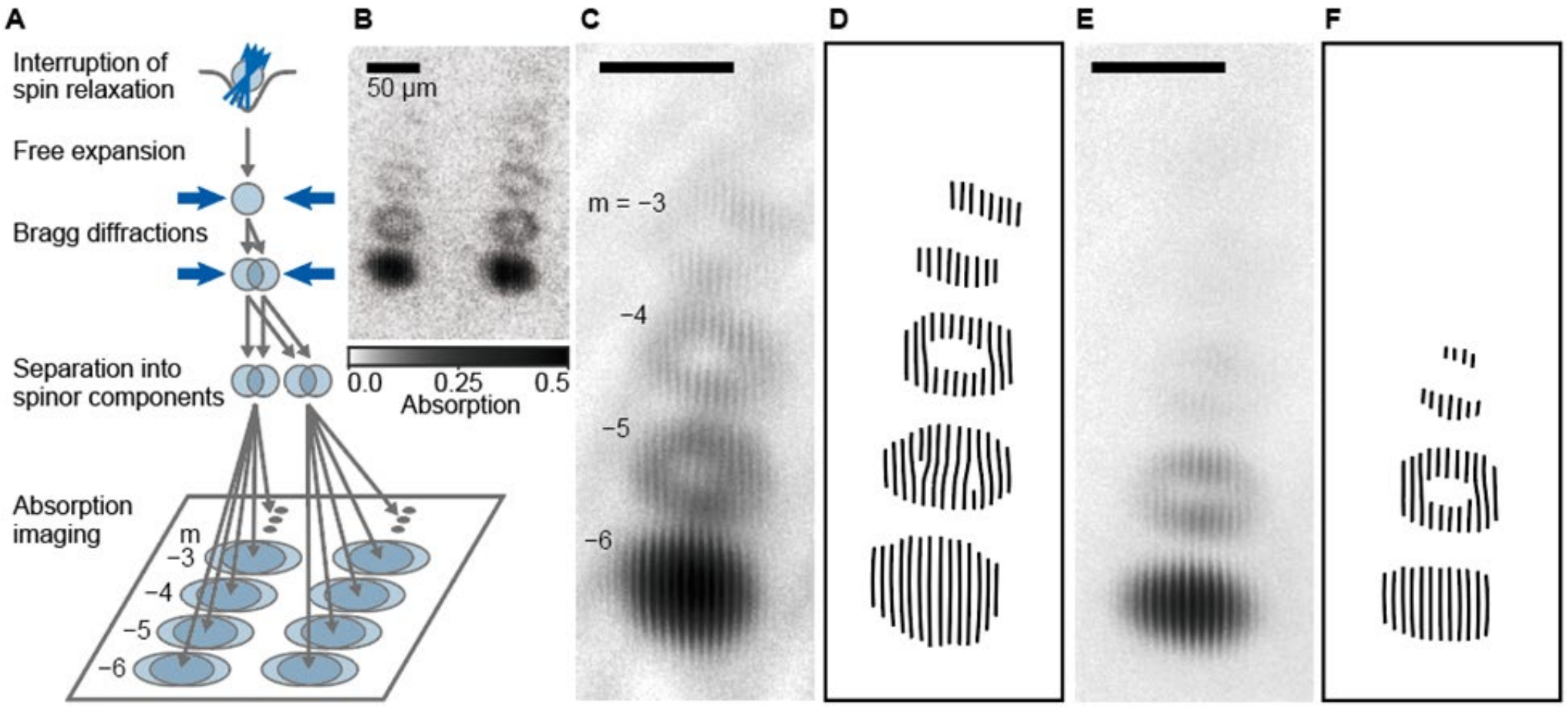

Fig. 3. Observation of mass circulation. (A) After a spin relaxation is interrupted by applying a magnetic field of 0.1 mT and released from the trap, two subsequent Bragg pulses diffract approximately half of the gas each, thereby constituting an open-type matter-wave interferometer. The atomic gases in the two output ports are spatially separated into spinor components and subsequently imaged. (B) Single-shot absorption image of the interferometer output for a gas held in a magnetic field of 2.6 nT for 5 ms. (C) Averaged image of 116 iterations with the same parameters as those in (B). (D) Sketch of the interference patterns in (C). (E) Averaged image of 171 iterations for a gas held in a magnetic field of 2.6 nT for 40 ms. (F) Sketch of the interference patterns in (E).

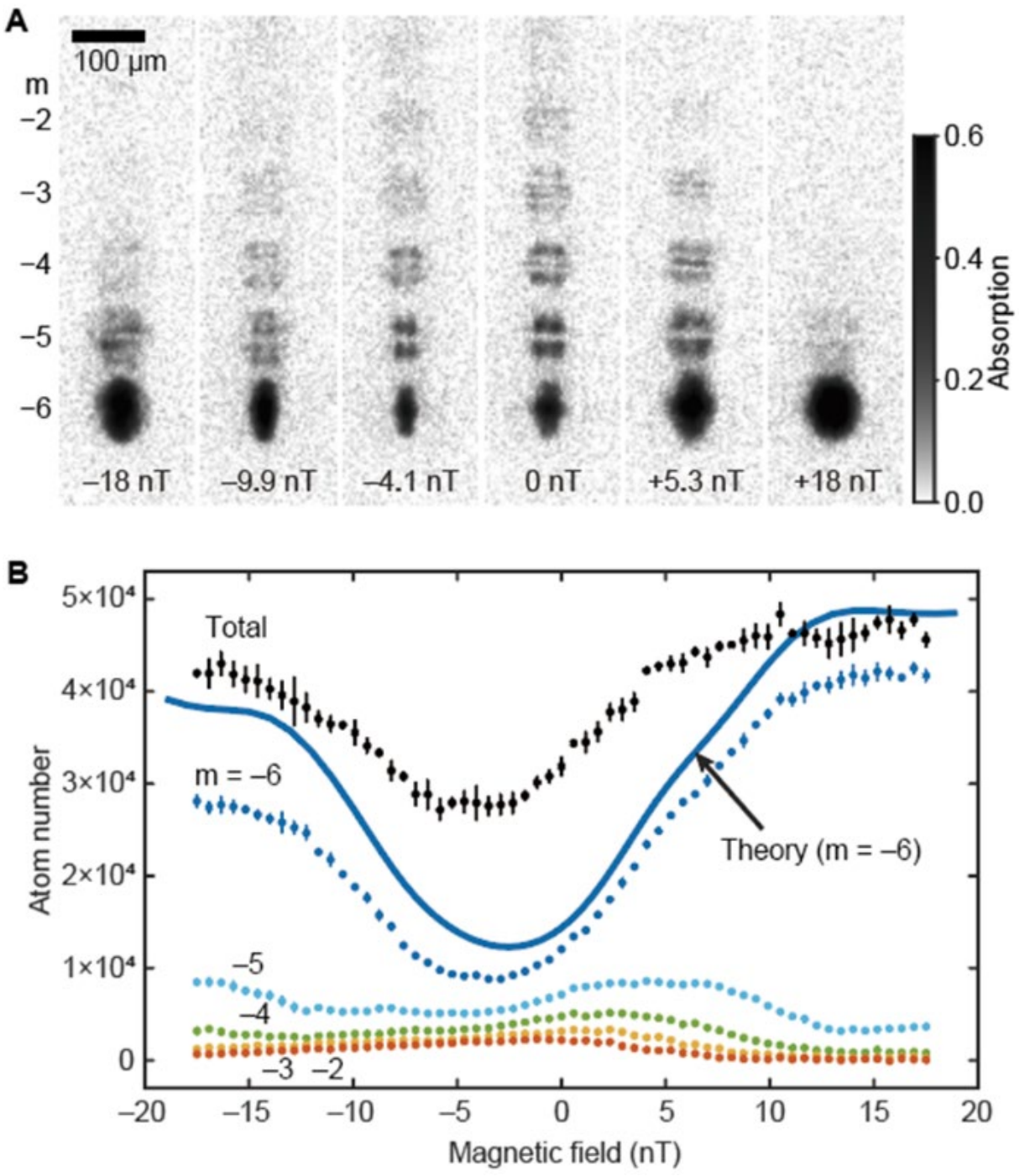

Fig. 4. Spin relaxations under various magnetic fields. (A) Single-shot absorption images of gases held in six magnetic fields for 5 ms. (B) Populations in the $m = -6, \ldots, -2$ spinor components after the gases were held in various magnetic fields for 5 ms (the $m \geq -1$ components are omitted for clarity). Each data point is the average of 4 experimental iterations. The error bar denotes the standard deviation. The magnetic fields shown may contain an offset error of up to 10 nT and random fluctuations of approximately 1 nT. The solid curve depicts the numerical results for the in-situ population in the $m = -6$ component.

## Resonant spin relaxation

In the aforementioned experiments, the spinor dynamics in atomic gases were initiated by a magnetic field of 2.6 nT. The dynamics are expected to vary under different magnetic fields, given that the external field determines the influence of the MDDI on the atomic spins (*37*). In the extreme case, spin relaxation does not occur, as we observed for a spin-polarized atomic gas held in a magnetic field of 1.0 µT. Moreover, it is predicted that the spin relaxation process will exhibit resonant enhancement when the magnetic field is reversed, whereby the highest energy spinor component is populated, and the Zeeman energy, spacing between magnetic sublevels, is equal to the rotational kinetic energy per atom (*21*).

To investigate the effect of a magnetic field on spinor dynamics, atomic gases were subjected to varying magnetic fields for a period of 5 ms, after which the spinor components were observed. The experimental procedure was identical to that previously utilized to observe spin relaxation, with the exception of the target magnetic field. The direction of the magnetic field was kept parallel to the initial vertical axis throughout the rapid ramps, from +1.0 µT to the desired value and then to +0.1 mT.

Typical single-shot images are shown in Fig. 4A. The population of each spinor component and the total population are shown in Fig. 4B. The dynamics under various magnetic fields were numerically simulated, and the resultant population in the $m = -6$ component is shown in Fig. 4B. Despite the absence of loss mechanisms in our model, the result of our numerical simulations remarkably resembles the experimental results. In Fig. 4B, both the experimental and theoretical results exhibit a dip in the population in the $m = -6$ component, with a minimum at a negative magnetic field. This offset can be interpreted as the resonant EdH effect. The difference between the estimated value of −1.5 nT, which is derived from the rotational kinetic energy (*38*, *39*), and the observed value of -3.5 nT can be attributed to magnetic dipole fields generated by the gas, which also contribute to the total magnetic field and consequently shift the resonance. The width of the dip can be regarded as the spread in the spatially distributed dipole field strength. The observed resonant enhancement supports that the spin relaxation is driven by the MDDI.

## Conclusion and Discussion

In conclusion, we observed spin relaxation accompanied by macroscopic circulation induced by the MDDI in a BEC, i.e., the EdH effect. As demonstrated in this study, the MDDI between atomic spins is fully responsible for the angular momentum transfer in a BEC, in which the microscopic and macroscopic torques are balanced. This is accomplished by transitions between the spinor components of the macroscopic wavefunction with the corresponding vorticity. While such balance in the MDDI can also be understood classically (*40*), it is important to note that in a BEC the transfer of angular momentum proceeds coherently. Although MDDI plays a minor role in the magnetism of rigid bodies (*41*), the essence of the EdH effect is the angular momentum conservation, which we have demonstrated by directly observing the phase winding of the macroscopic wavefunction via matter-wave interferometry. Furthermore, we demonstrated that the EdH effect in a BEC emerges spectacularly via dynamical symmetry breaking of spherical symmetry.

This phenomenon represents a hallmark coherent dynamics of a spinor-dipolar BEC. The combined effect of the MDDI, the kinetic energy, and the spin-dependent short-range interactions in a spinor-dipolar BEC gives rise to a variety of ground-state phases, some of which spontaneously break the chiral symmetry while generating a chirality-dependent local circulating flow that does not contribute to the net circulation. The observation of the EdH effect opens the door to study such ground-state phases with spin textures and mass circulation (*24*, *25*, *42*, 43). Moreover, the Barnett effect, magnetization induced by mechanical rotation, in a BEC can be investigated by preparing a spinor BEC with an imprinted vortex (44, 45).

**Acknowledgment**

This work was supported by JSPS Japan KAKENHI Grant No. JP16K13856, No. JP20J21364, No. JP22H01152 and No. JP24K00557, the Murata Science Foundation, and the Research Foundation for Opto-Science and Technology.

## Appendix A: Magnetic field environment

The magnetic field environment plays a crucial role in this work. A stable magnetic field is realized by a fully passive measure using magnetic shields, as in Refs. (*47, 48*). Our magnetic shield (Ohtama Co., Ltd.) comprises three mu-metal layers with small openings, thus providing the necessary optical access for the experiments, as shown in Fig. S1. The desired weak magnetic field environment is maintained through a daily demagnetization procedure. The shield is equipped with a demagnetizing coil comprising 28 turns of generic electric wire neatly and evenly laid on the mu-metal layers in a toroidal coil configuration. The alternating current in the coil exhibits a quadratic decay in amplitude, declining from 1.2 to 0 A over the seven-minute duration of the demagnetization procedure. The frequency is maintained at 2 Hz throughout. The alternating current for the procedure was generated by our homemade current source. Its current output is AC coupled to the coil through an ordinary isolation transformer (rated 1000 VA at 50 Hz) because we could not reach residual magnetic fields below 1 nT with an offset current of approximately 10 μA in our current source output. Prior to incorporation of the shield in the vacuum apparatus, the residual magnetic field in the shield was measured with a commercial optical pumping magnetometer (QuSpin, Inc. QZFM Gen-3): the bias field was less than 1.0 nT, and the magnetic field gradient was less than 10 nT/m.

The magnetic field applied to the gas at the center of the shield was controlled by three pairs of Helmholtz-like coils and one pair of anti-Helmholtz-like coils. The coils are situated inside the shield and firmly secured by a polycarbonate fixture, thereby eliminating eddy currents. The electric currents in the coils were controlled by homemade bipolar current sources, the design of which was based on those described in Refs. (*49*, *50*). The current source can provide a current up to 1.0 A with a time constant of 30 μs without exhibiting overshoot. Control signals can be dynamically attenuated by 20 dB of the original value within an experimental run, thereby extending the dynamic range for controlling the current. The magnetic fields generated by the coils in the shield were calibrated by radiofrequency spectroscopy (*51*) of the $F = 6, m = -6 - F = 6, m = -5$ transition of Eu atoms. Although the coils can be employed to cancel the field remaining in the shield at the location of the gas, the method of nulling the total field with an accuracy of 0.1 nT (a Larmor frequency of 1.6 Hz) is not trivial: the MDDI in an atomic gas dominates the Zeeman interaction under such weak magnetic fields. However, the spatial distributions of the spinor components are very sensitive to magnetic fields, as shown in Fig. 4. This can be utilized in nulling the magnetic field as described below.

Throughout the experiments depicted in Fig. 4, the weak magnetic fields applied to the gases were parallel to the initial polarization axis of the atomic spins. When the gas was subjected to a magnetic field inclined with respect to the initial polarization axis for a duration of 5 ms, deformation of the lateral segmentation in the middle of the $m = -5$ component was observed (Fig. S2). By tuning the horizontal magnetic field so that the lateral segmentation appears horizontal, the radial component in the total magnetic field can be reliably nulled with an accuracy on the order of 0.1 nT. Furthermore, applying a weak magnetic field for an extended duration of 20 ms increases the sensitivity of the spatial distribution to the total magnetic field because the magnetic dipolar field generated by the gas becomes weaker because of population loss. As the weak field is tuned toward negative, at a certain value, a clear segmentation in the $m = -5$ component is no longer visible. At this value, the direction of the total magnetic field applied to the gas is determined essentially by fluctuations in the residual magnetic field. Consequently, we presume that the total magnetic field is practically zero at the value. Given that the longitudinal magnetic field offset is determined in this manner, we present a relatively large systematic error of 1 nT for the magnetic field applied to the gas in this work.

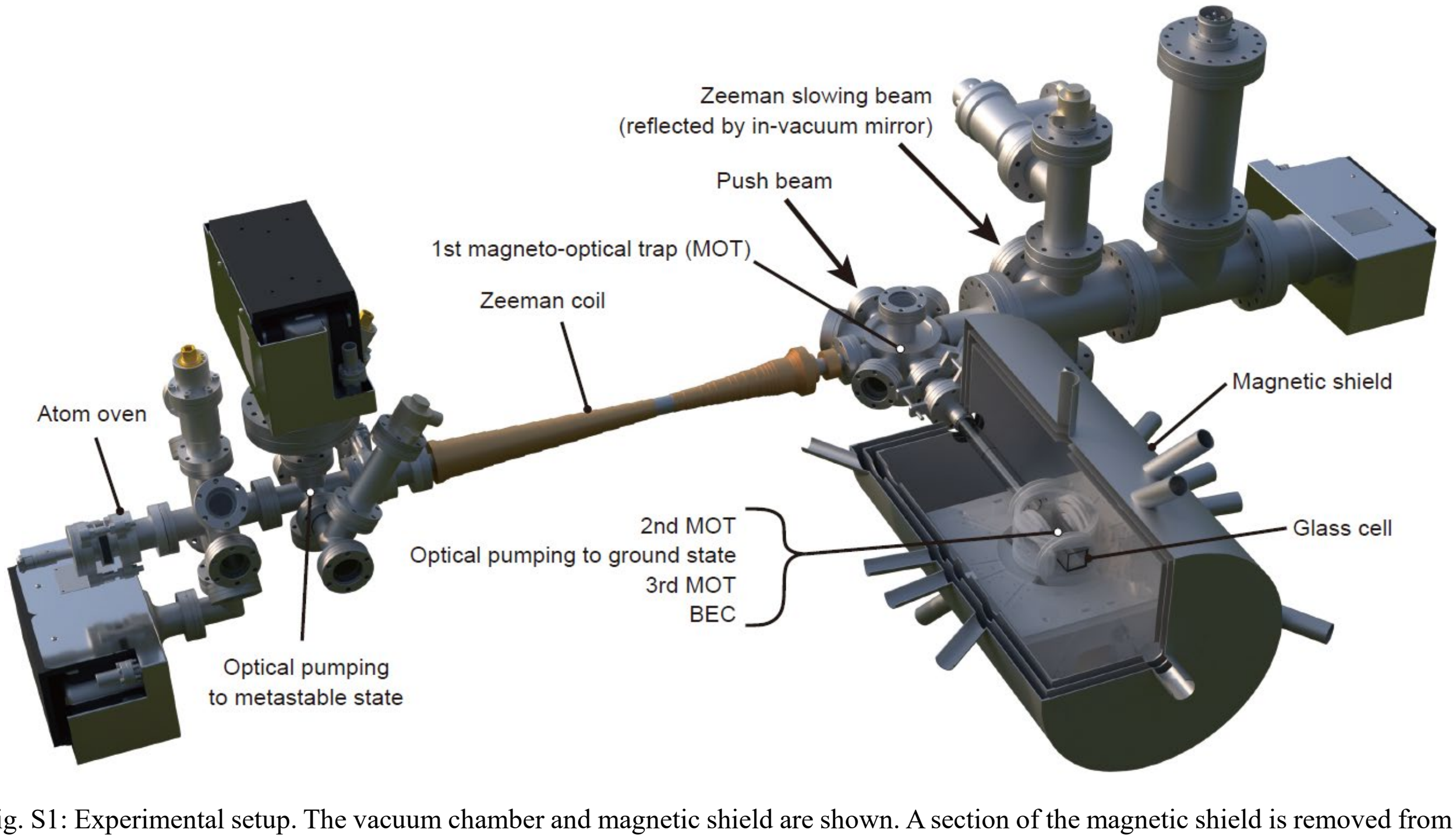

Fig. S1: Experimental setup. The vacuum chamber and magnetic shield are shown. A section of the magnetic shield is removed from the view to allow visualization of the glass cell and internal coil fixture.

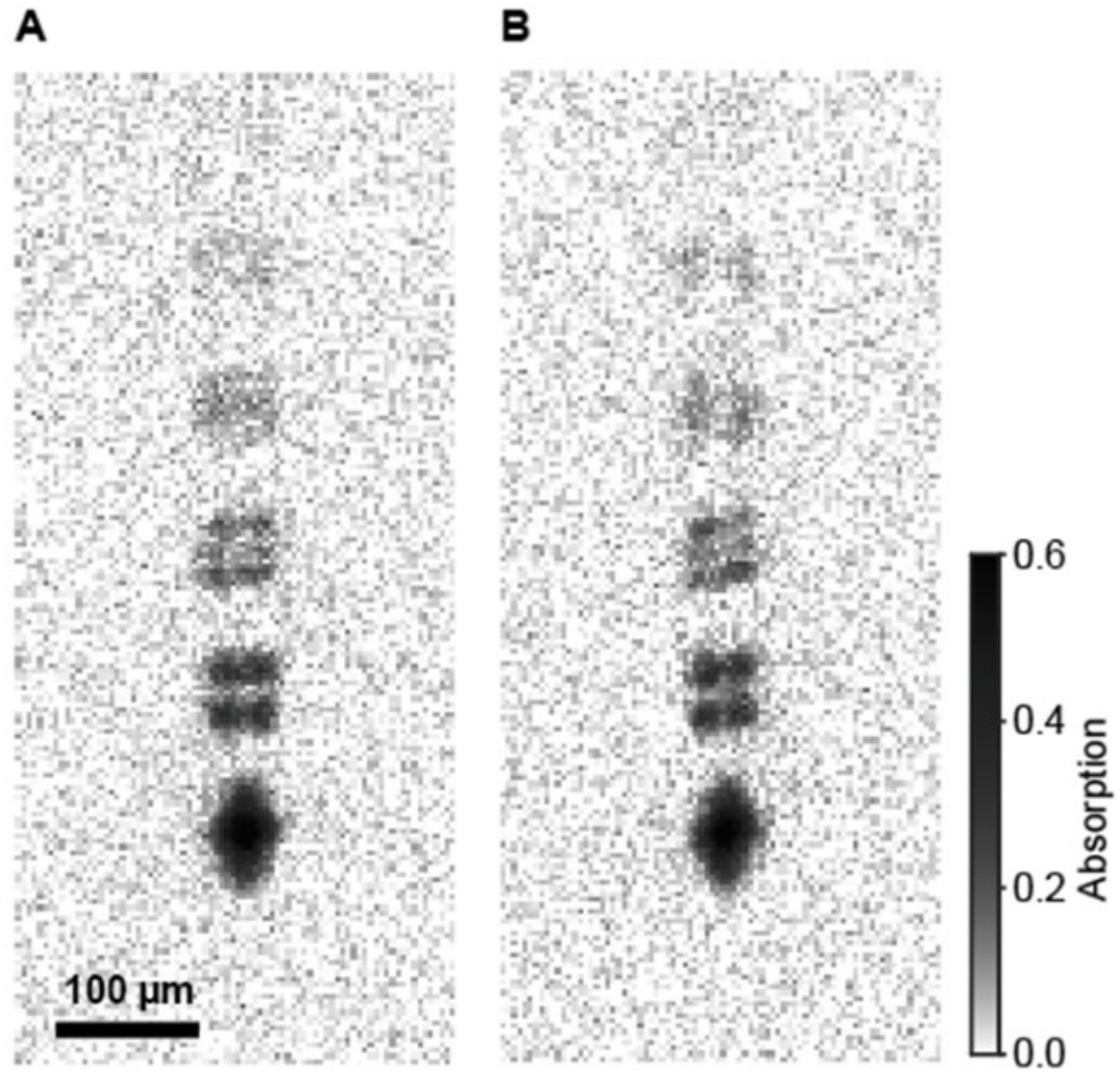

Fig. S2: Effect of horizontal residual magnetic field on the spatial profiles of the spinor components. (A) The spin-polarized atomic gas is held in a weak magnetic field of 2.6 nT oriented upright for a duration of 5 ms and subsequently imaged following the same procedure as that in Fig. 1E. (B) A lateral magnetic field of approximately 0.1 nT is added to the weak magnetic field.

The magnetic shield contains several optical elements in addition to the aforementioned coils: two lenses to focus laser beams for the optical dipole trap, one objective lens for horizontal imaging, and one microscope objective lens for high-resolution vertical imaging. The polyetheretherketone (PEEK) lens barrel of the microscope objective is supported by four glass rods penetrating the shields, allowing fine adjustment of the objective lens position from outside the shield. In addition, each fixture within the magnetic shield is made of polycarbonate, which should prevent disturbance of the magnetic field environment by eddy currents and magnetization of the fixture materials.

## Appendix B: Preparation of the Eu BEC

The preparation procedure is based on the latest report (*26*). Eu atoms supplied from a 770-K oven were briefly laser-cooled in the first magneto-optical trap (MOT) outside the magnetic shield. Laser cooling was performed using the $a^{10}D_{13/2}$ $F$=9 – $z^{10}F_{15/2}$ $F'$=10 transition at a wavelength of 583 nm with a natural linewidth of 8.2 MHz available in the electronic metastable state ($a^{10}D_{13/2}$ $F$=9) because the electronic ground state ($a^{8}S_{7/2}$) lacks broad cyclic optical transitions practically appropriate for cooling gases above room temperature (*52*). The first MOT was operated with a quadrupole magnetic field (40 mT/m axial gradient) oriented vertically and cooling laser beams (6.6 mW each, with a 22-mm waist diameter) red-detuned by 18 MHz from resonance. The cooling laser beam contained a sideband resonantly addressing the $F$=8 – $F'$=9 transition to optically pump back atoms in the $F$=8 hyperfine state. The atoms in the first MOT were continuously outcoupled at the right angle by a collimated push beam, resulting in a cold atomic beam of presumably 1 mK. The push beam (0.11 mW, with a 1.0-mm waist diameter, linearly polarized along the horizontal direction) addresses the same transition utilized in the first MOT at the same detuning but does not contain sidebands for repumping. The cold atomic beam was directed into the magnetic shield and was captured by the second MOT, which is situated at the center of the magnetic shield. This resulted in a horizontal transfer of 0.48 m of atoms into the shield. This configuration resembles the double MOT described in Ref. (*53*). The slight vertical inclination of the push beam compensates for gravitational acceleration during the flight between the MOTs. The second MOT was operated with a quadrupole magnetic field oriented approximately perpendicular to the cold atomic beam in the horizontal plane. The other parameters for the second MOT were similar to those for the first MOT. The transfer efficiency between the two MOTs was evaluated from the loading rates and reached 26(3)%. The captured atoms were continuously optically pumped back to the electronic ground state ($a^{8}S_{7/2}$) and were recaptured by the third MOT, which shares the quadrupole magnetic field with the second MOT. In the third MOT we utilize the $a^{8}S_{7/2}$ $F$=6 – $z^{10}P_{9/2}$ $F'$=7 transition at a wavelength of 687 nm with a natural linewidth of 97 kHz (*54*). Finally, the captured atoms were loaded into an optical dipole trap produced by a horizontal laser beam and evaporatively cooled at the intersection of the horizontal laser beam and a vertically inclined laser beam. In the evaporative cooling process, the applied magnetic field was maintained at 0.3 mT oriented vertically and determined the lowest energy spin state of $F$=6, $m$=−6, in which atoms reach quantum degeneracy.

To allow the gas to circulate smoothly in the trap potential, the condensed atoms were transferred to a round-shaped optical trap before spinor dynamics were initiated in the low magnetic field. The round trap was realized by intersecting two horizontal laser beams: one to which atoms are loaded and another with a waist radius of 50 μm. The optical axis of the second one was placed 10 μm above the optical axis of the first one, resulting in a highly symmetric trap potential. The transfer from the initial trap to the round trap was adiabatically performed by linearly ramping the power of the two laser beams simultaneously for 0.2 s. The gas confined in the round trap contains a negligible thermal component and $5 \times 10^4$ atoms in a polarized spin state. The magnetic field was decreased in a stepwise manner from 0.3 mT to the desired value via 1 μT, thus ensuring predictable and controlled entry into the weak magnetic field region. The resultant atomic gas was imaged using an absorption imaging technique (*55*) with circularly polarized probe light at a wavelength of 460 nm resonantly addressing the $a^8S_{7/2}$ $F$=6 – $y^{10}P_{9/2}$ $F'$=7 transition.

### Appendix C: Matter-wave interferometry

The experimental matter-wave interferometry procedure is shown in Fig. 3A. Atomic gas released from the trap is irradiated by a pair of horizontally counter-propagating laser beams, which constitute a light crystal and induce Bragg diffraction of atoms (*30*). The frequency difference between the two lasers is tuned to couple two momentum states of each atom separated by two-photon recoils in momentum. Momentary exposure of the gas to the diffraction light pair results in the coherent creation of a copy of the gas with an additional momentum that eventually displaces the copy from the original. Following an interval, another momentary exposure creates a copy of the original with an additional momentum and, conversely, a copy of the first copy with the original momentum. The two momentum states resulting from the two consecutive pulses constitute two output ports of the interferometer. Each output port contains a pair of atomic gases that are displaced between the two pulses. The two gases within each port expand and eventually spatially overlap, thereby exhibiting interference fringes that reflect the phase difference within the initial atomic gas (*33*, *34*, *36*).

The experimental parameters are identical to those previously employed to observe spin relaxations up to the point at which gas is released from the trap. Following free expansion of 2.1 ms, the atomic gas, which had been subjected to a weak field, was irradiated by two pulses separated by an interval of 0.9 ms. The two counter-propagating laser beams with a wavelength of $\lambda$ = 460 nm address the same atomic transition as the probe light but are detuned below the $F = 6 - F' = 7$ resonance by 0.8 GHz, which is set much greater than the natural linewidth of 27 MHz to suppress spontaneous emission. The temporal amplitude envelope of each diffraction pulse is Gaussian with a width of 7.5 μs, ensuring uniform diffraction of the entire gas sample with a momentum width of $0.3\hbar k$, where $\hbar k = 2\pi\hbar/\lambda$ is the momentum of a photon with a wavelength of $\lambda$ (*31*). The peak intensity of each laser beam is tuned to result in approximately equal populations in the initial and the coherently diffracted momentum states following each irradiation event. The diffraction light is horizontally polarized and oriented perpendicular to the applied magnetic field. Prior to imaging, the atomic gas was separated into spinor components along the horizontal direction perpendicular to the diffraction light beams (Fig. 3A). This enabled the observation of the phase distributions of the spinor components in a single view through our non-magnetic high-resolution microscope objective with a resolution of 0.86 μm, situated below the atomic gas. The resulting absorption image, acquired after a time of flight of 12 ms, revealed atomic gases from the two interferometer output ports, with a displacement of 110 μm. Additionally, the pair of gases within each port, displaced by 10 μm, exhibited spatial interference with a typical fringe spacing of 3.1 μm.

### Appendix D: Spin-dependent atom loss

As briefly mentioned in the text, spin relaxation and atom loss are notably suppressed under a sufficiently strong magnetic field. When the gas was held in a strong magnetic field of 1 μT for an extended duration of 16 s, spin relaxation was not observed, as expected since the atomic spins are pinned to the external field. The loss dynamics of the population were dominated by a three-body loss process with an initial loss rate of 0.55 /s, corresponding to a loss of 2% over 40 ms. Therefore, the loss process inherent to such spin-polarized gases give a negligible contribution to the rapid loss of 38% over 40 ms in Fig. 2B. This suggests that the loss mechanism highly depends on the spin states of colliding atoms.

When the application of the magnetic field of 0.1 mT fo interrupting the spin relaxation was suspended until 2 ms after the release when the gas had sufficiently expanded, the loss was reduced to 26% of the total populations over 40 ms. The Zeeman energy in the weak field and MDDI were on the order of 10 nK, which is insufficient to cause atoms to escape from the trap (1.3 μK deep). Thus, the atoms were likely to be lost in three-body collisions involving atoms not in the $m = -6$ component, whereby the molecular binding energy is converted into kinetic energy. The remaining 12% of the losses could be attributed to collisions of atoms during the initial 2 ms of free expansion under the strong magnetic field of 0.1 mT, thereby converting the Zeeman energy into kinetic energy. Notably, interrupting the spin relaxation with weaker magnetic fields yielded images with blurry spatial structures, indicating the incomplete interruption. This is likely due to the reduced ramp rate in the interrupting magnetic field. We consider that a trade-off exists between population loss and image quality in the magnetic field for spin projection, with 0.1 mT representing the optimal solution.

## Appendix E: Theoretical model

We numerically investigated the dynamics of the 13-component spinor wavefunction $\Psi$ using the time-dependent Gross–Pitaevskii (GP) equation: $\mathrm{i}\hbar\frac{\partial \Psi_m}{\partial t} = \frac{\delta H}{\delta \Psi_m^*}$, where $\hbar$ is the reduced Planck constant. The model and calculation method follow those presented in Ref. (*19*). The mean-field Hamiltonian $H$ consists of the single-particle Hamiltonian $H_0$, the MDDI $H_{\mathrm{dd}}$, and the short-range (van der Waals) interaction $H_{\mathrm{sr}}$, namely, $H = H_0 + H_{\mathrm{dd}} + H_{\mathrm{sr}}$. The single-particle Hamiltonian comprises the kinetic energy and the confining potential energy, which are spin independent, as well as the Zeeman energy in the external magnetic field. A spin-6 Bose gas has 7 independent parameters characterizing its short-range interactions, but currently, only one parameter is experimentally known for spin-6 $^{151}$Eu BEC: the scattering length of two polarized atoms $a_{12} = 110a_{\mathrm{B}}$ ($a_{\mathrm{B}}$ denotes the Bohr radius). For simplicity, we assumed the short-range interactions consisting of only the density-density and spin-spin interactions, i.e., $H_{\mathrm{sr}} = \frac{1}{2}\int[c_0 n(\mathbf{r})^2 + c_1 \mathbf{f}(\mathbf{r})^2]d^3\mathbf{r}$, where $n(\mathbf{r}) = \Psi^\dagger(\mathbf{r})\Psi(\mathbf{r})$ and $\mathbf{f}(\mathbf{r}) = \Psi^\dagger(\mathbf{r})\mathbf{F}\Psi(\mathbf{r})$ are the number density and the spin density, respectively, with $\mathbf{F}$ denoting the vector of spin $F$=6 matrices. The coefficients $c_0$ and $c_1$ are constrained to give the same mean-field energy for the spin-polarized gas as $c_0 + F^2 c_1 = 4\pi\hbar^2 a_{12}/M$, where $M$ is the mass of the Eu atom. The ratio $c_1/c_0$ was determined to best reproduce the observed spatial profiles. The MDDI Hamiltonian can be expressed as $H_{\mathrm{dd}} = \frac{g_F\mu_{\mathrm{B}}}{2}\int \mathbf{B}_{\mathrm{dd}}(\mathbf{r})\cdot\mathbf{f}(\mathbf{r})\mathrm{d}^3\mathbf{r}$, where $\mathbf{B}_{\mathrm{dd}}(\mathbf{r}) = \frac{g_F\mu_{\mathrm{B}}\mu_0}{4\pi}\int\left\{\frac{\mathbf{f}(\mathbf{r}')}{|\mathbf{r}-\mathbf{r}'|^3} - \frac{3[\mathbf{f}(\mathbf{r}')\cdot(\mathbf{r}-\mathbf{r}')](\mathbf{r}-\mathbf{r}')}{|\mathbf{r}-\mathbf{r}'|^5}\right\}\mathrm{d}^3\mathbf{r}'$ is the magnetic dipole field created by the gas with $g_F$ and $\mu_B$ being the Landé g-factor and the Bohr magneton, respectively. When the external magnetic field is much larger than the dipole field, we can use the time-averaged MDDI over the Larmor precession period (*37*).

We prepared the initially spin-polarized state using the imaginary-time propagation method (*56*) with the time-averaged MDDI and calculated the real-time dynamics under a weak external field with the full MDDI. In the simulation for the imaging process, we again switched to the time-averaged MDDI when the magnetic field was ramped to 0.1 mT for interrupting the spin relaxation and expanded the gas for 2.7 ms. We use the Crank–Nicolson scheme to calculate the time evolution (*57*) up to this part. Because the atomic density becomes low enough after the free expansion of 2.7 ms, we solve the remaining time evolution in momentum space while neglecting the interactions. In all the calculations, we did not consider atomic losses.

The spin flip in the initial stage of the dynamics is well understood by describing the MDDI in terms of the ladder operators $F^\pm \equiv F^x \pm \mathrm{i}F^y$, where the MDDI term in the GP equation reads $\frac{1}{2}g_F\mu_{\mathrm{B}}(2\,B_{\mathrm{dd}}^z(\mathbf{r})F^z + B_{\mathrm{dd}}^+(\mathbf{r})F^- + B_{\mathrm{dd}}^-(\mathbf{r})F^+)\,\Psi(\mathbf{r})$, with $B_{\mathrm{dd}}^\pm(\mathbf{r}) \equiv B_{\mathrm{dd}}^x(\mathbf{r}) \pm \mathrm{i}B_{\mathrm{dd}}^y(\mathbf{r})$. This expansion yields the selection rule for the MDDI: atoms can change their magnetic sublevel by up to one. When the atomic spins are fully polarized in the $m = -6$ component, i.e., $\Psi_m(\mathbf{r}) = \delta_{m,-6}\sqrt{n(\mathbf{r})}$, only the MDDI is responsible for population transfer to the $m = -5$ component: $\mathrm{i}\hbar\frac{\partial}{\partial t}\Psi_{-5}(\mathbf{r}) = \frac{1}{2}\,g_F\mu_{\mathrm{B}}F_{-5,-6}^+ B_{\mathrm{dd}}^-(\mathbf{r})\sqrt{n(\mathbf{r})}$. Considering the circular symmetry around the *z*-axis and the inversion symmetry about the *z*=0 plane of the initial polarized BEC in cylindrical coordinates $(\rho, \phi, z)$, the spatial dependence of $B_{\mathrm{dd}}^-(\rho,\phi,z)$ is given by $z\rho e^{-\mathrm{i}\phi}$ times a positive function of $\rho$ and $|z|$. This in turn yields the spatial dependence of $\Psi_{-5}(\mathbf{r})$ with a phase winding of $-2\pi$, as described in the text. Subsequently, atoms in the $m = -5$ component can transfer to the $m = -4$ component, thereby acquiring an additional spatial phase winding of $-2\pi$. This process continues as long as $B_{\mathrm{dd}}^-(\mathbf{r})$ does not change significantly.

Notably, population transfer between the spinor components via the short-range interactions also preserves a phase winding of $2\pi\times(-6-m)$ in $\Psi_m(\mathbf{r})$. Following population transfer driven by the MDDI in the initial stage, the difference of $2\pi$ between the phase windings in $\Psi_m(\mathbf{r})$ and $\Psi_{m+1}(\mathbf{r})$ results in phase windings of $\pm 2\pi$ in the spin density components $f^\pm(\mathbf{r}) \equiv f^x(\mathbf{r}) \pm \mathrm{i}f^y(\mathbf{r})$. Therefore, the spin-exchanging part of the short-range interaction, namely, $c_1(f^+(\mathbf{r})F^- + f^-(\mathbf{r})F^+)\Psi(\mathbf{r})$ in the GP equation, preserves the phase windings in the spinor components. The change in the phase winding is a consequence of the local spin-gauge symmetry and local spin exchange in a gas possessing spin texture. No net orbital angular momentum is transferred in this process. Notably, in the initial fully polarized state, the transverse spin density and, accordingly, the spin-exchanging part of the short-range interaction are absent. Thus, the short-range interaction does not contribute to the initial stage of the dynamics.

The short-range interaction dependence of the dynamics appears in the $m = -4$ component. In our simulation, spin-independent ($c_1 \sim 0$) and ferromagnetic ($c_1 < 0$) short-range interactions resulted in only two rings in the component. The results shown in Fig. 1, F and G, feature three rings in the $m = -4$ component, and they were obtained for $c_0 = 2\pi\hbar^2 a_{12}/M$ and $c_1 = (1/$ $18)\pi\hbar^2 a_{12}/M$, which best reproduce the spatial profile in the experimental results. This suggests an antiferromagnetic nature of the short-range interactions in spin-6 $^{151}$Eu gas with high magnetization.